\documentclass[12pt]{elsart}

\usepackage{graphicx}
\usepackage{subfigure}
\usepackage{amsmath}
\usepackage{amssymb}
\usepackage{cite}
\usepackage{endfloat}

\journal{Physics Letters A}

\begin{document}

\begin{frontmatter}

%opening
\title{Absence of long-range order in a general spin-$S$ kagome lattice Ising antiferromagnet}
\author{M. Semjan},
\author{M. \v{Z}ukovi\v{c}\corauthref{cor}}
\ead{milan.zukovic@upjs.sk}
\address{Institute of Physics, Faculty of Science, P.J. \v{S}af\'arik University\\ Park Angelinum 9, 041 54 Ko\v{s}ice, Slovakia}
\corauth[cor]{Corresponding author.}

\maketitle

\begin{abstract}
The possibility of the emergence of some kind of long-range ordering (LRO) due to the increase of multiplicity of the local degrees of freedom (spin value $S$) is studied in an Ising antiferromagnet on a kagome lattice (IAKL) by Monte Carlo simulation. In particular, the critical exponent of the spin correlation function, obtained from a finite-size scaling analysis, is evaluated for various values of $S$, including $S=\infty$, with the goal to determine whether there exists some threshold value of the spin $S_C$ above which the system would show true or quasi-LRO, similar to a related model on a triangular lattice (IATL). It is found that, unlike in the IATL case, the IAKL model remains disordered for any spin value and any finite temperature.
\end{abstract}

\begin{keyword}
Ising antiferromagnet \sep kagome lattice \sep general spin \sep geometrical frustration \sep long-range order 

%\PACS 05.50.+q \sep 64.60.De \sep 75.10.Hk \sep 75.30.Kz \sep 75.50.Ee \sep 75.50.Lk

\end{keyword}

\end{frontmatter}

\section{Introduction}

Geometrical frustration in spin systems prevents simultaneous minimization of all microscopic interactions and weakens the system's tendency to form an ordered state at low or even zero temperature. The conceptually simplest example of such a system is a spin $S=1/2$ Ising antiferromagnet on a triangular lattice (IATL), which has been shown to exhibit no long-range ordering (LRO) at any finite temperatures~\cite{wann50,wann73,hout50}. The ground state ($T=0$) is characterized by a finite residual entropy (0.3231$k_B$) and a power-law decaying correlation function with the exponent $\eta=1/2$~\cite{step70}.

One way to alleviate the effects of frustration and thus to induce order is by increasing multiplicity of the local degrees of freedom, which can be achieved by increasing the magnitude of the spin variable $S$. The effect of the spin magnitude on critical properties of IATL has attracted considerable attention and numerous results on this topic have been reported in a series of papers~\cite{hori91,hori92,hori93,naga93,netz93,naga94,yama95,lipo95,zeng97}. A phenomenological theory lead to a conclusion that in the limit of $S \to \infty$ the ground state shows (partial) antiferromagnetic LRO with two sublattices fully ordered and the third one disordered. This scenario was further supported by Monte Carlo simulations. They found that at low temperatures with the increasing $S$ the value of the correlation function exponent decreases from $\eta=1/2$ for $S=1/2$ down to zero for $S$ exceeding some threshold value $S_C$, suggesting that for sufficiently large values of $S$ LRO exists at zero temperature and a phase transition occurs at a finite temperature.

The spin $S=1/2$ quantum Heisenberg antiferromagnet on a triangular lattice (HATL) was initially believed to show liquid-like ground state without magnetic LRO~\cite{ande73,faze74}. Nevertheless, a series of later studies lead to a conclusion that for any $S$, including the extreme quantum limit of $S=1/2$, the ground state displays a semi-classical three-sublattice N\'{e}el LRO~(see, e.g., Refs.~\cite{joli89,bern92,chub94,capr99,whit07,zhit13,gotz16}), albeit very fragile due to the interplay between quantum fluctuations and strong frustration with the sublattice magnetization drastically diminished~\cite{capr99,whit07}.

A number of investigations have also been carried out for the Heisenberg antiferromagnet on a kagome lattice (HAKL), providing a very strong numerical evidence that the ground state of the model is a quantum spin liquid with no magnetic LRO for spin $S=1/2$~\cite{sach92,chub92,henl95,yan11,gotz11,depe12,jian12,iqba15} as well as for the $S=1$ case~\cite{gotz11,liu15,chan15,nish15,li15}. Nevertheless, magnetic LRO can be stabilized by increasing the spin quantum number to $S>1$~\cite{gotz11,cher14,gotz15,oitm16,liu16} even though it melts at any finite temperature~\cite{liu16,mull18}. 

Considering the attention paid to the existence of LRO in the general spin-$S$ IATL and HATL models on the triangular lattice as well as the HAKL model on the kagome lattice, it is surprising that no study of this kind has been done yet for a general spin-$S$ Ising antiferromagnet on the kagome lattice (IAKL). For $S=1/2$ IAKL displays even higher geometrical frustration than IATL resulting in no LRO at any temperature including zero temperature limit~\cite{syoz51} and extremely large value of the residual entropy (0.5018$k_B$)~\cite{kano53}. The ground state is characterized by a short-range order (spin liquid state), with two spins parallel and the third one antiparallel in each triangle. However, there is no ordering among the triangles, which leads to a massive degeneracy~\cite{harr92,chalk92,reim93,delf93}. 

In the present study we focus on the question whether some kind of ordering can be induced by increasing multiplicity of the local degrees of freedom even in such a superfrustrated spin system. In particular, we consider the spin-$S$ IAKL model, with $S$ increasing up to infinity (continuous spin), and study the character of the spin-correlation function by Monte Carlo simulations. 

\section{Model and method}
The Hamiltonian of the spin-$S$ IAKL model is given by
\begin{equation}
    \mathcal{H} = -J\sum_{\langle i,j\rangle }{\sigma_i\sigma_j}, 
\label{eq:H}
\end{equation}
where $J<0$ is the exchange interaction, $\sigma_i = S_i/S$, where $S_i$ denotes the Ising spin on the $i$-th site allowed to take values $-S$, $-S+1$, ..., $S-1$, $S$, and the summation runs over all nearest neighbors. For $S\rightarrow\infty$ the spin variables $\sigma_i$ are allowed to take continuous values from the interval $\langle -1, 1\rangle$.

The model is studied by employing the Monte Carlo (MC) method with the standard Metropolis algorithm and focusing on the low-temperature region. We execute extensive MC simulation runs on kagome lattices with the linear sizes $L$ = 36, 60, 84, 128 and 160 (corresponding to the total number of spins $N=3L^2$) with the periodic boundary conditions, for several increasing values of the spin $S$ = 1/2, 1, 3/2, 5/2, 4, and $\infty$. The simulations start from random states at a relatively high temperature $T_{max}=3.5$ and continue to lower temperatures with gradually decreasing step $\Delta T$ down to $T_{min}=0.003$, which is close to the ground-state conditions. The final state at the previous temperature is used as the initial state for the next temperature. For averaging in the equilibrium state we take over $4\times 10^6$ MC sweeps after discarding the initial $10^6$ sweeps, which are used for thermalization. Throughout the paper we set $J=-1$ and $k_B=1$.

\begin{figure}[t!]
    \center
    \includegraphics[width = 0.48\textwidth]{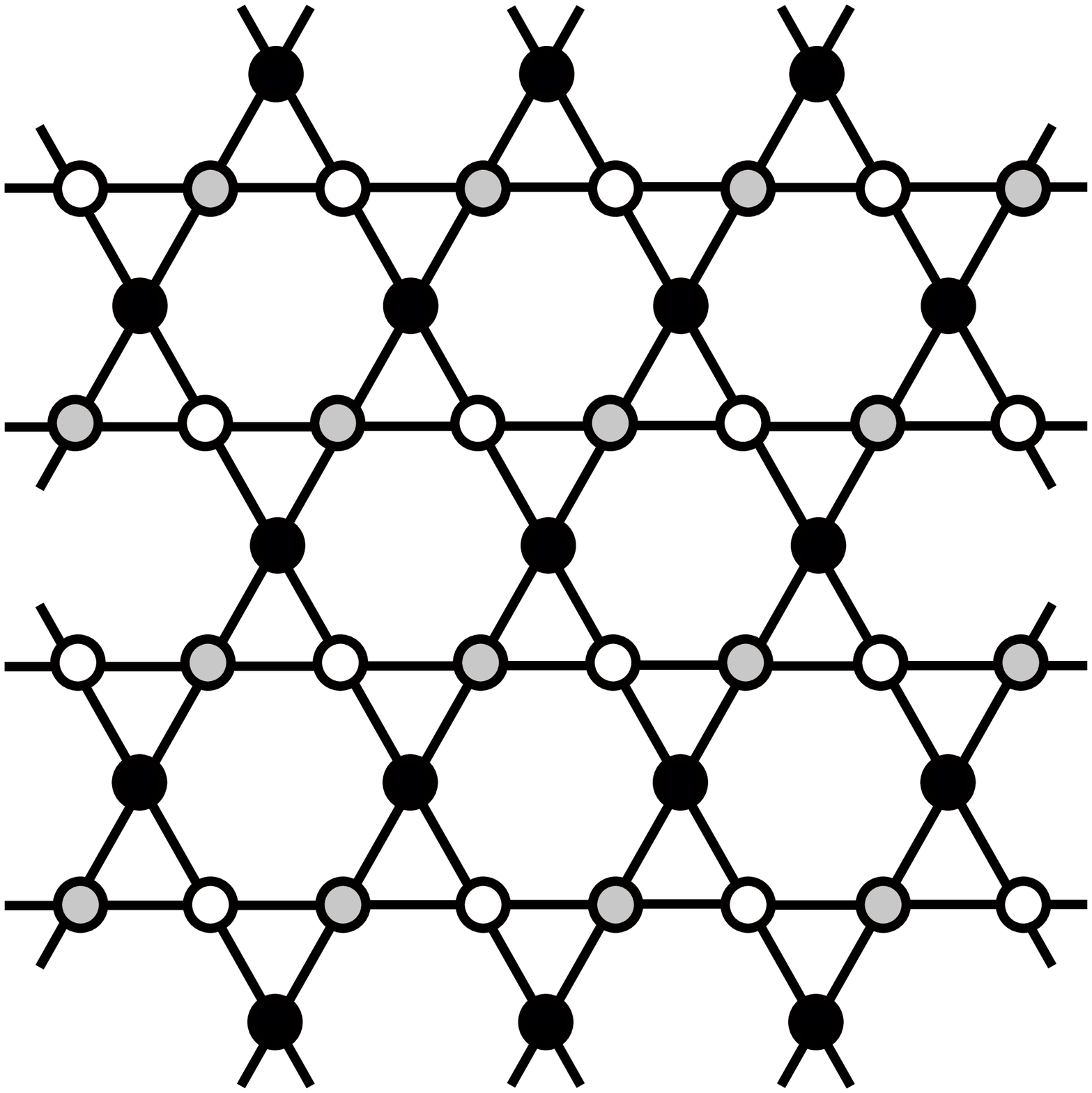}
    \caption{Kagome lattice partitioned into three interpenetrating sublattices, A (shaded circles), B (open circles), and C (filled circles), with the ${\mathbf q}=0$ structure.}
    \label{fig:kagome_q_0}
\end{figure}

We calculate the internal energy per spin $e=\left\langle \mathcal{H} \right\rangle/N$ and from its fluctuations the specific heat per spin, defined as 
\begin{equation}
c=\frac{\langle \mathcal{H}^{2} \rangle - \langle\mathcal{H} \rangle^{2}}{NT^{2}}.
\label{eq:c}
\end{equation}
We consider the kagome lattice consisting of three interpenetrating sublattices A, B and C with the so-called ``${\mathbf q}=0$'' structure, as shown in Fig.~\ref{fig:kagome_q_0}. If there is any sublattice LRO (true one or just quasi-LRO), it should be reflected in the behavior of sublattice magnetizations (sublattice order parameters), which can be calculated by summing all the spins in the respective sublattices, i.e.,
\begin{equation}
	m_{\alpha}=\left\langle M_{\alpha} \right\rangle/N = \left\langle \sum_{i\in \alpha}\sigma_i \right\rangle/N,\qquad \alpha = {\rm A,~B,~C}.
\label{eq:mi}
\end{equation}
If the system displays the quasi-LRO with the power-law decaying spin correlation function
\begin{equation}
	\left\langle S_{i}S_{j} \right\rangle \propto r_{ij}^{-\eta},
\label{eq:CF}
\end{equation}
where $\eta$ is the critical exponent, the sublattice order parameters, $m_{\alpha}$, should scale with the lattice size as~\cite{chal86}
\begin{equation}
	m_{\alpha}(L) \propto L^{-\eta/2}.
\label{eq:mi_FSS}
\end{equation}
Alternatively, the quantity
\begin{equation}
	Y = \left\langle M_{\rm A}^2+ M_{\rm B}^2+M_{\rm C}^2\right\rangle/N,
\label{eq:Y}
\end{equation}
related to the magnetic susceptibility, should scale as~\cite{miya86,naga94}
\begin{equation}
	Y(L) \propto L^{2-\eta}.
\label{eq:Y_FSS}
\end{equation}
True magnetic LRO can be detected if $\eta$ goes to zero, while the disordered phase with exponentially decaying correlation function will be characterized by the value of $\eta=2$.

\section{Results and conclusions}

\begin{figure}[t!]
    \center
		\subfigure{\includegraphics[width = 0.35\textwidth]{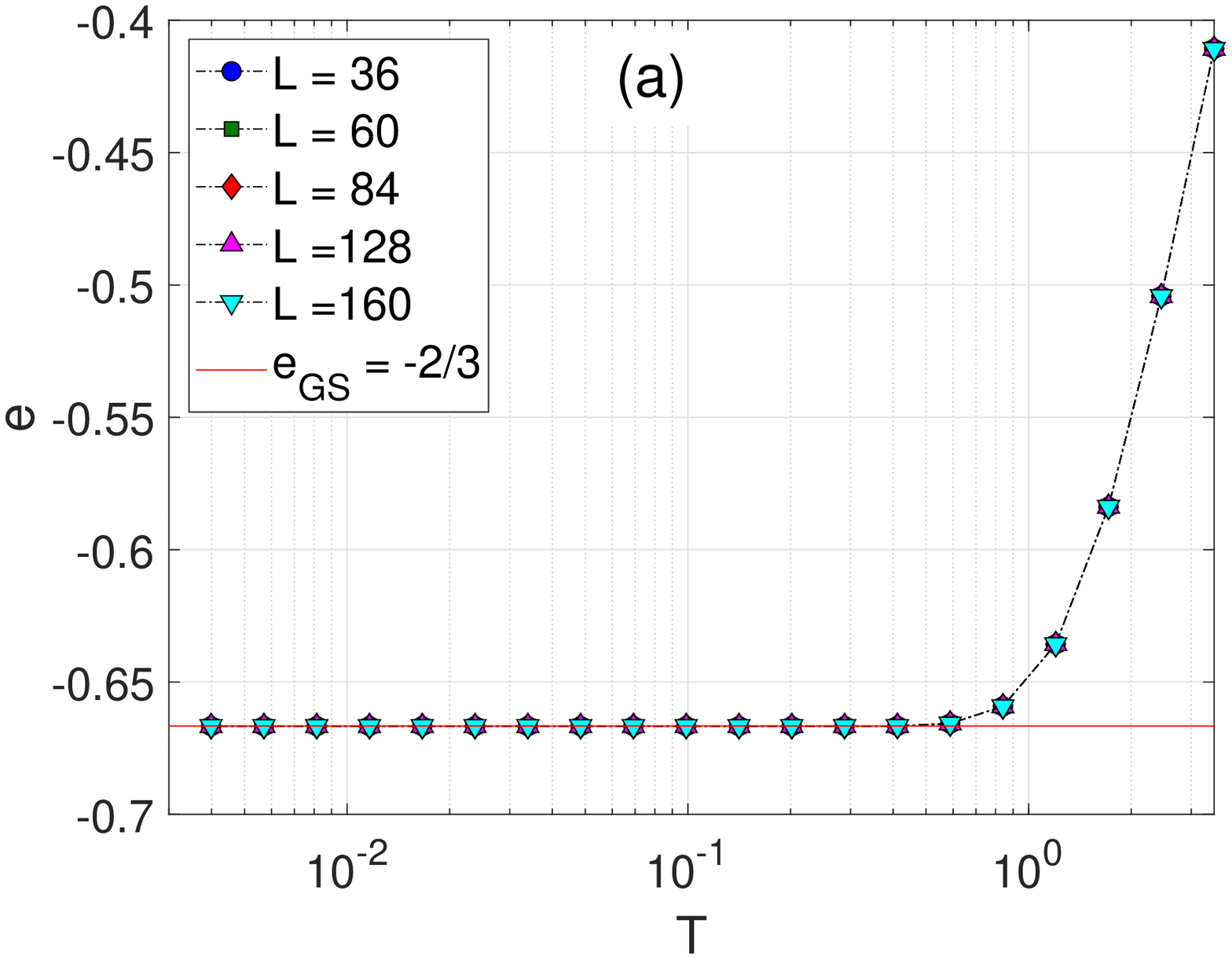}\label{fig:e_s_0_5}}\hspace{-5mm}
    \subfigure{\includegraphics[width = 0.35\textwidth]{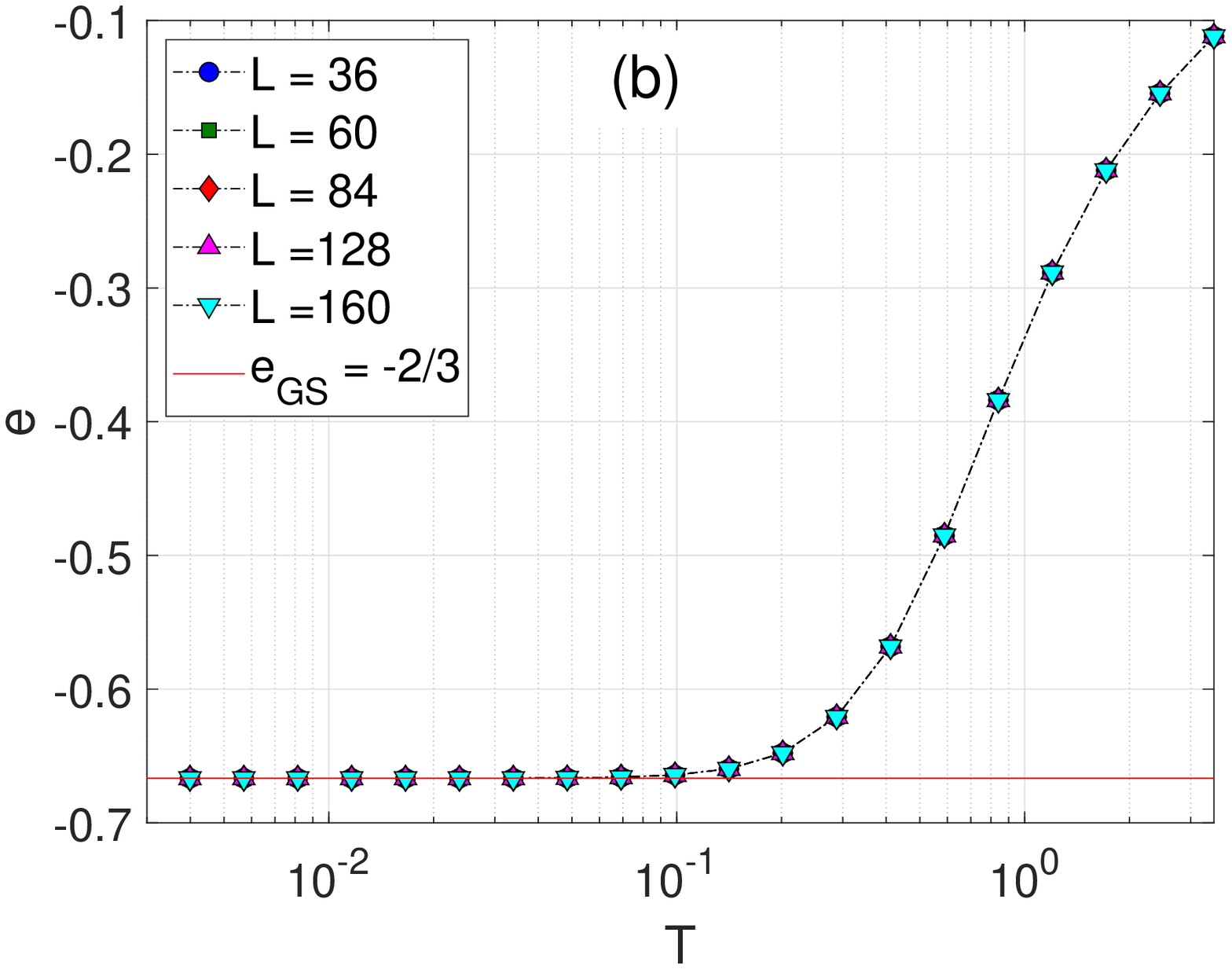}\label{fig:e_s_2_5}}\hspace{-5mm}
		\subfigure{\includegraphics[width = 0.35\textwidth]{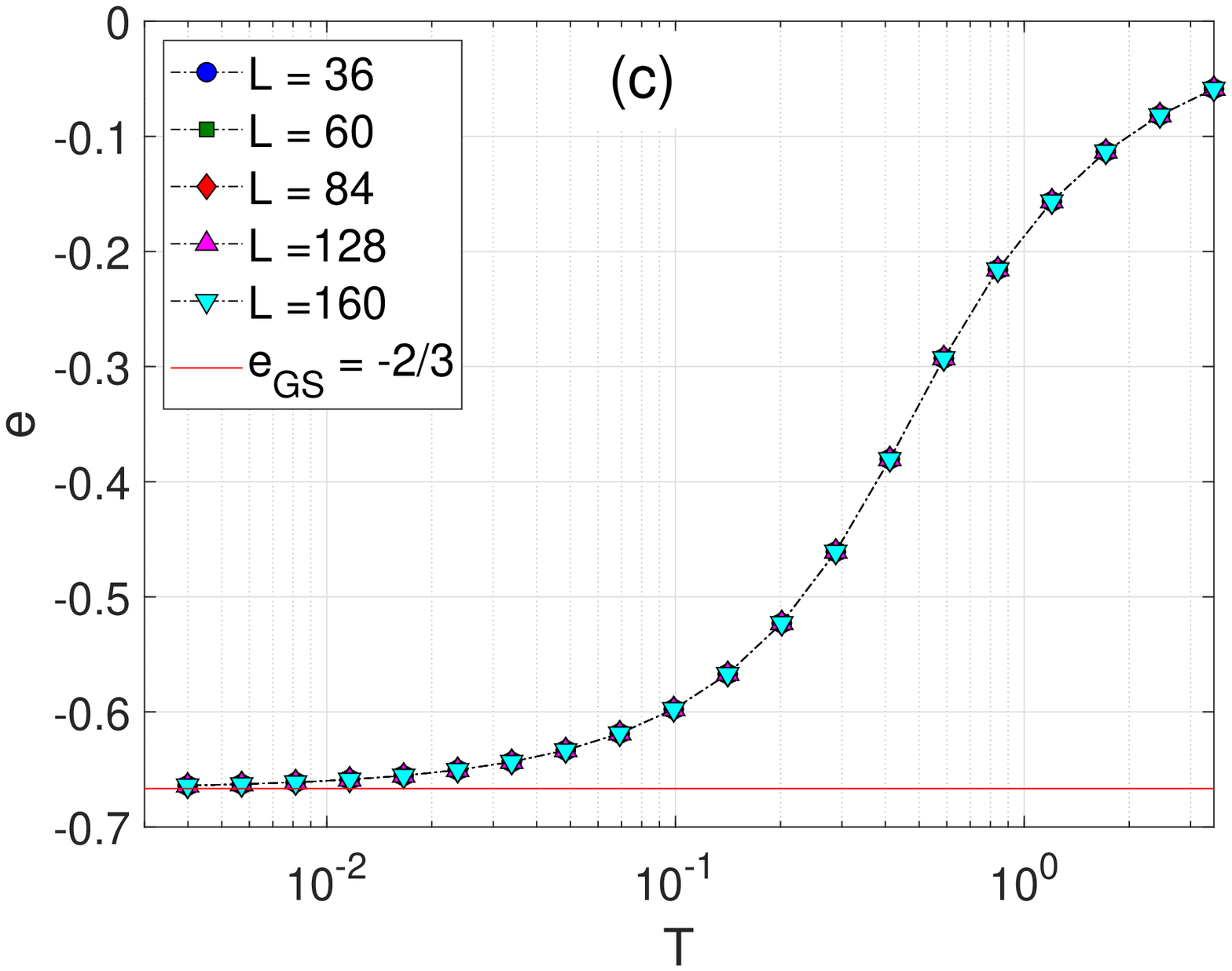}\label{fig:e_s_inf}}
		\subfigure{\includegraphics[width = 0.35\textwidth]{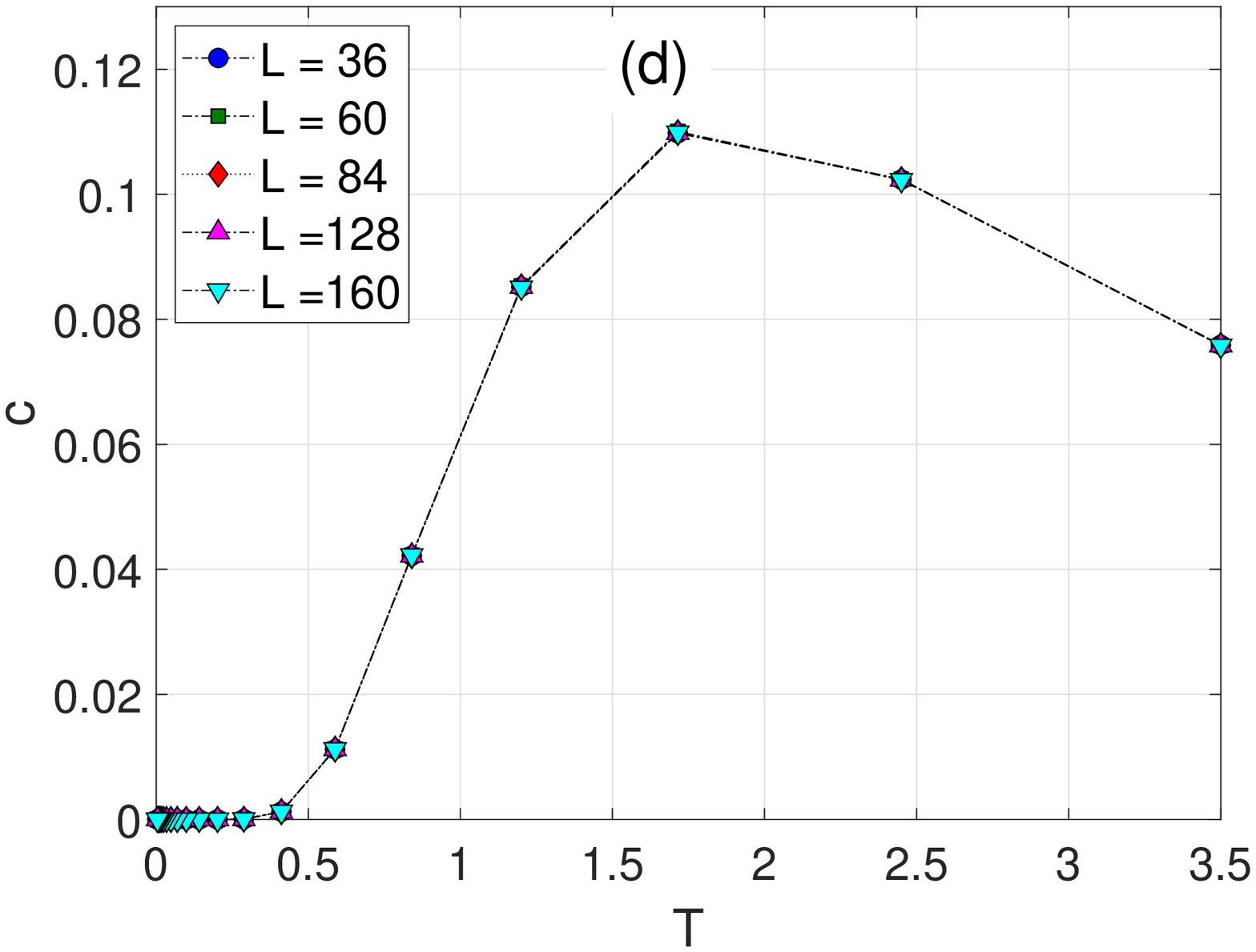}\label{fig:c_s_0_5}}\hspace{-5mm}
    \subfigure{\includegraphics[width = 0.35\textwidth]{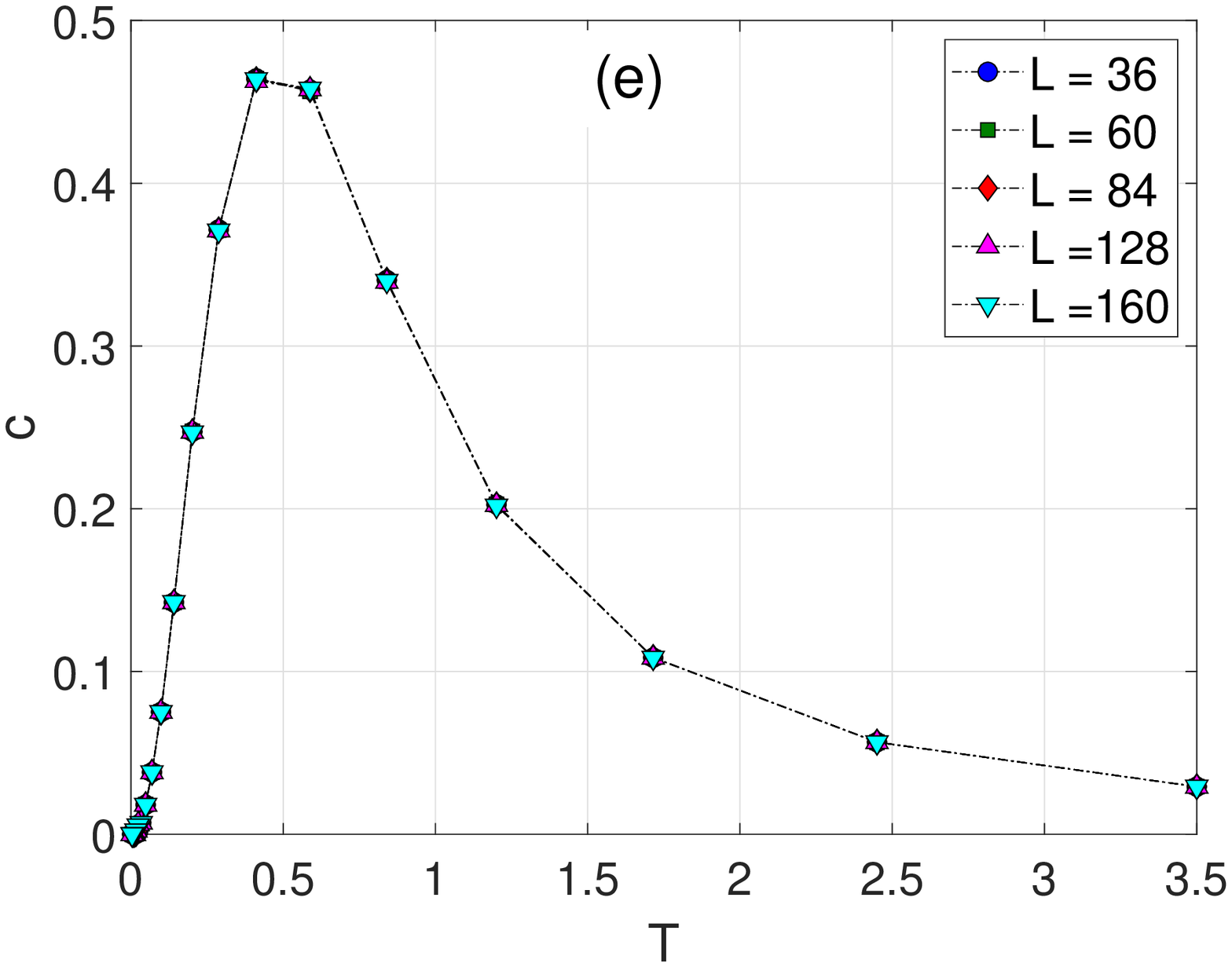}\label{fig:c_s_2_5}}\hspace{-5mm}
		\subfigure{\includegraphics[width = 0.35\textwidth]{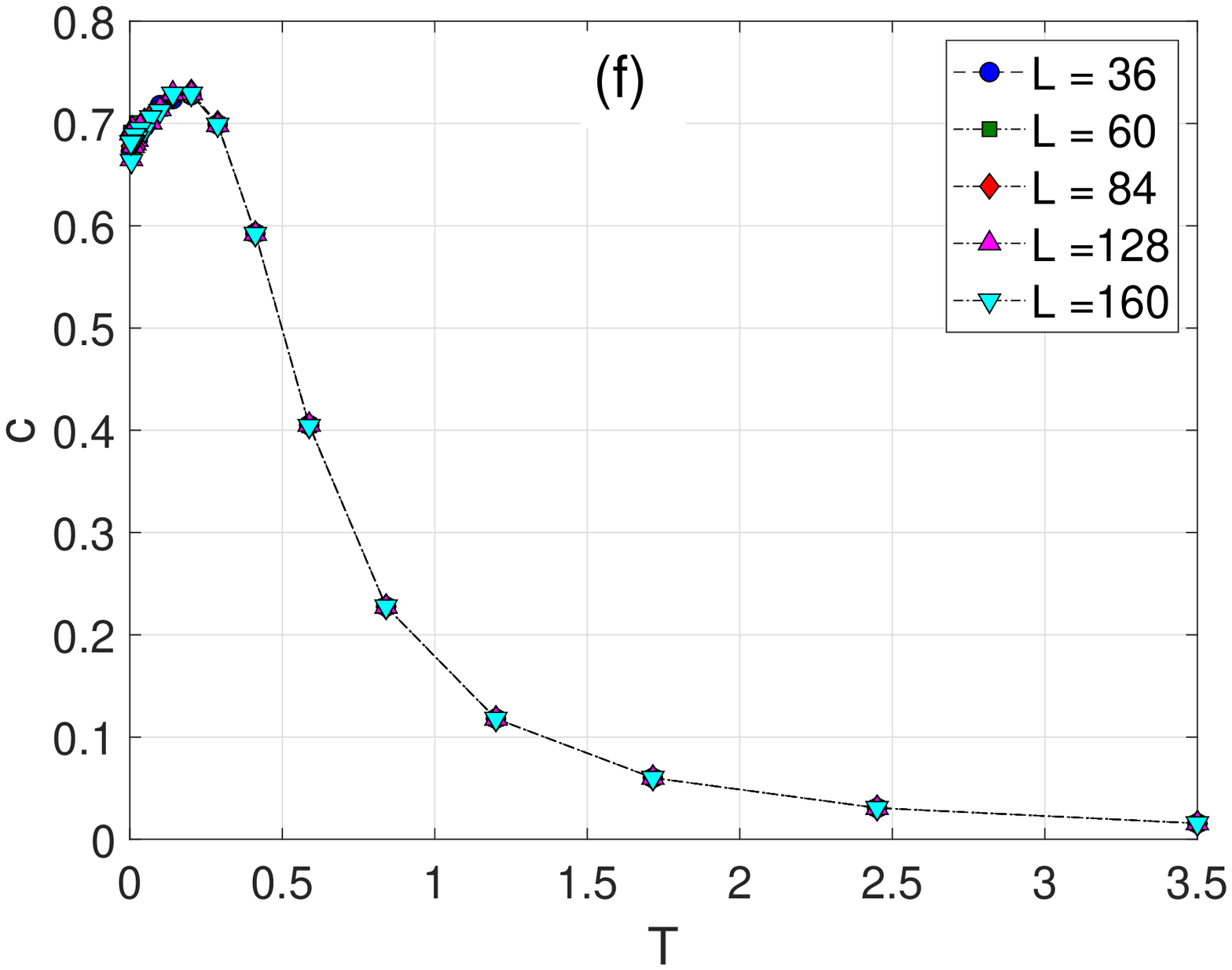}\label{fig:c_s_inf}}
    \subfigure{\includegraphics[width = 0.35\textwidth]{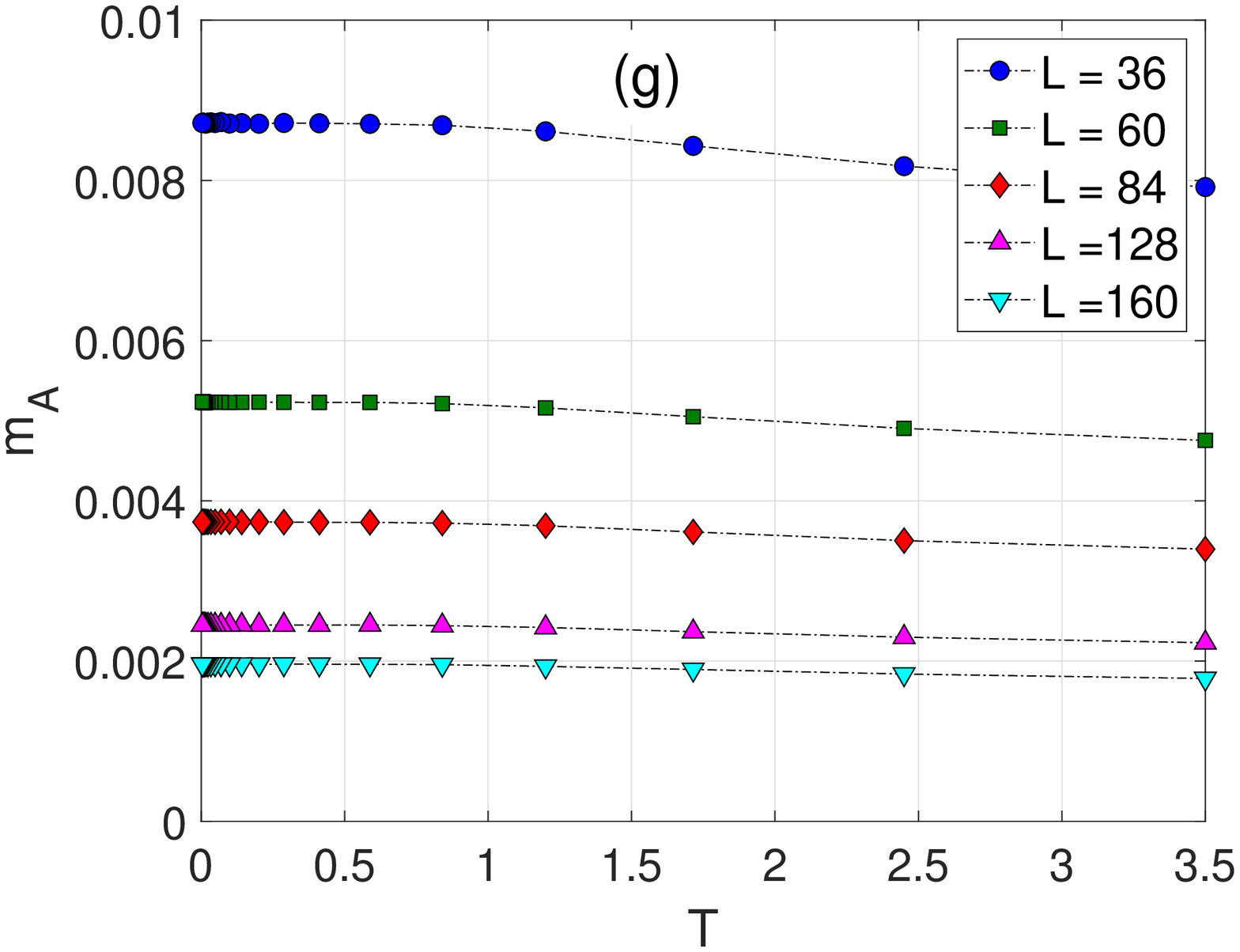}\label{fig:ma_s_0_5}}\hspace{-5mm}
    \subfigure{\includegraphics[width = 0.35\textwidth]{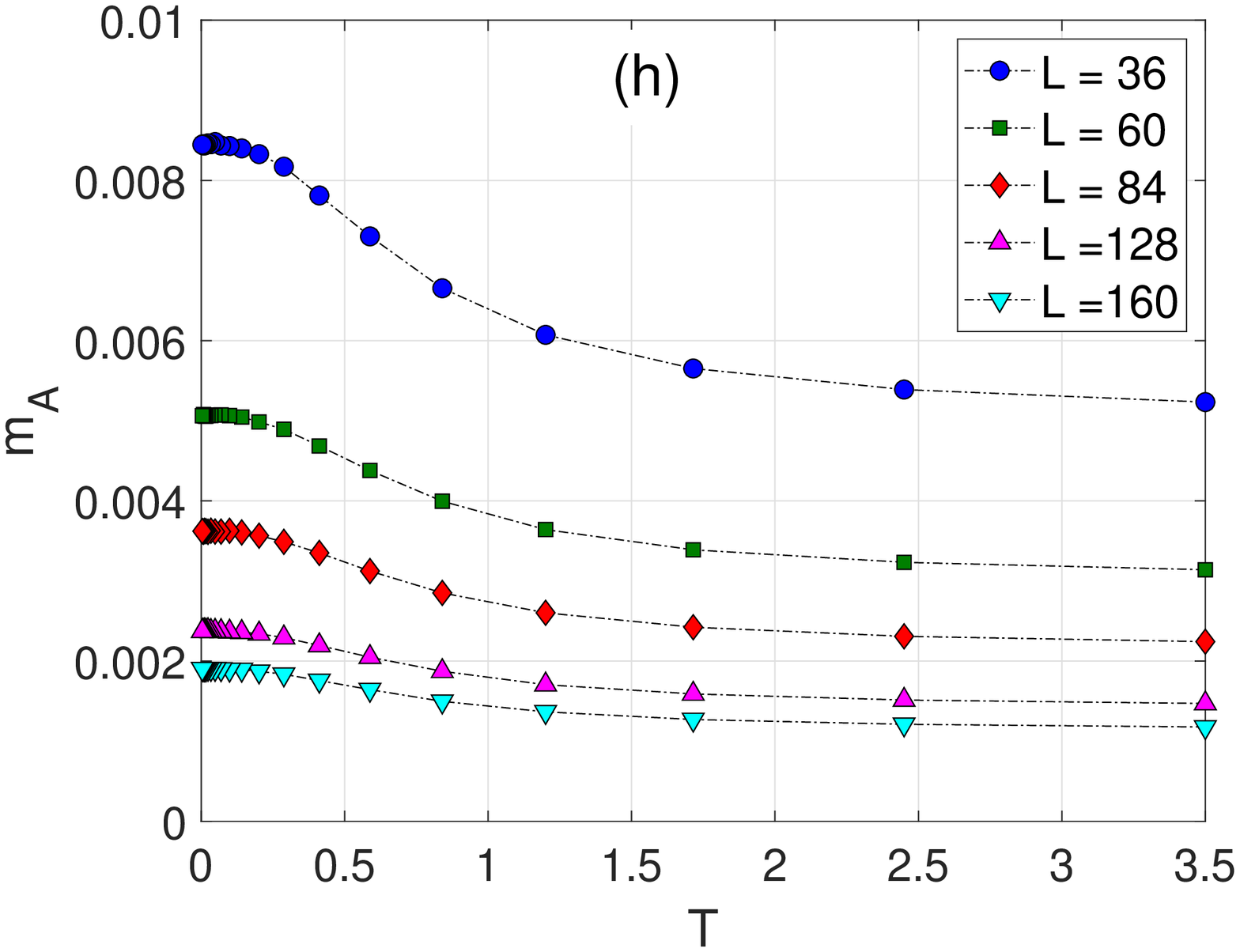}\label{fig:ma_s_2_5}}\hspace{-5mm}
		\subfigure{\includegraphics[width = 0.35\textwidth]{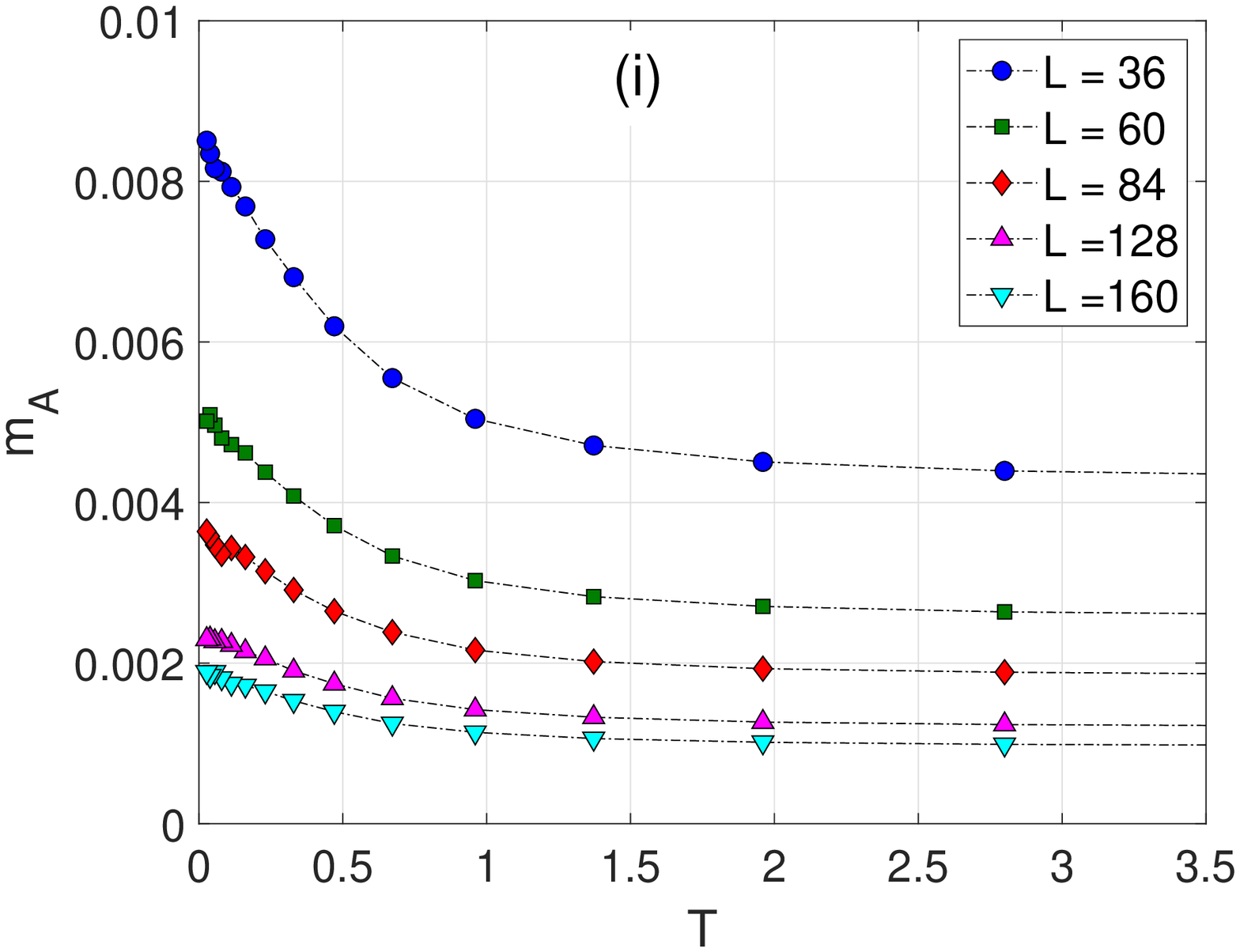}\label{fig:ma_s_inf}}
		\subfigure{\includegraphics[width = 0.35\textwidth]{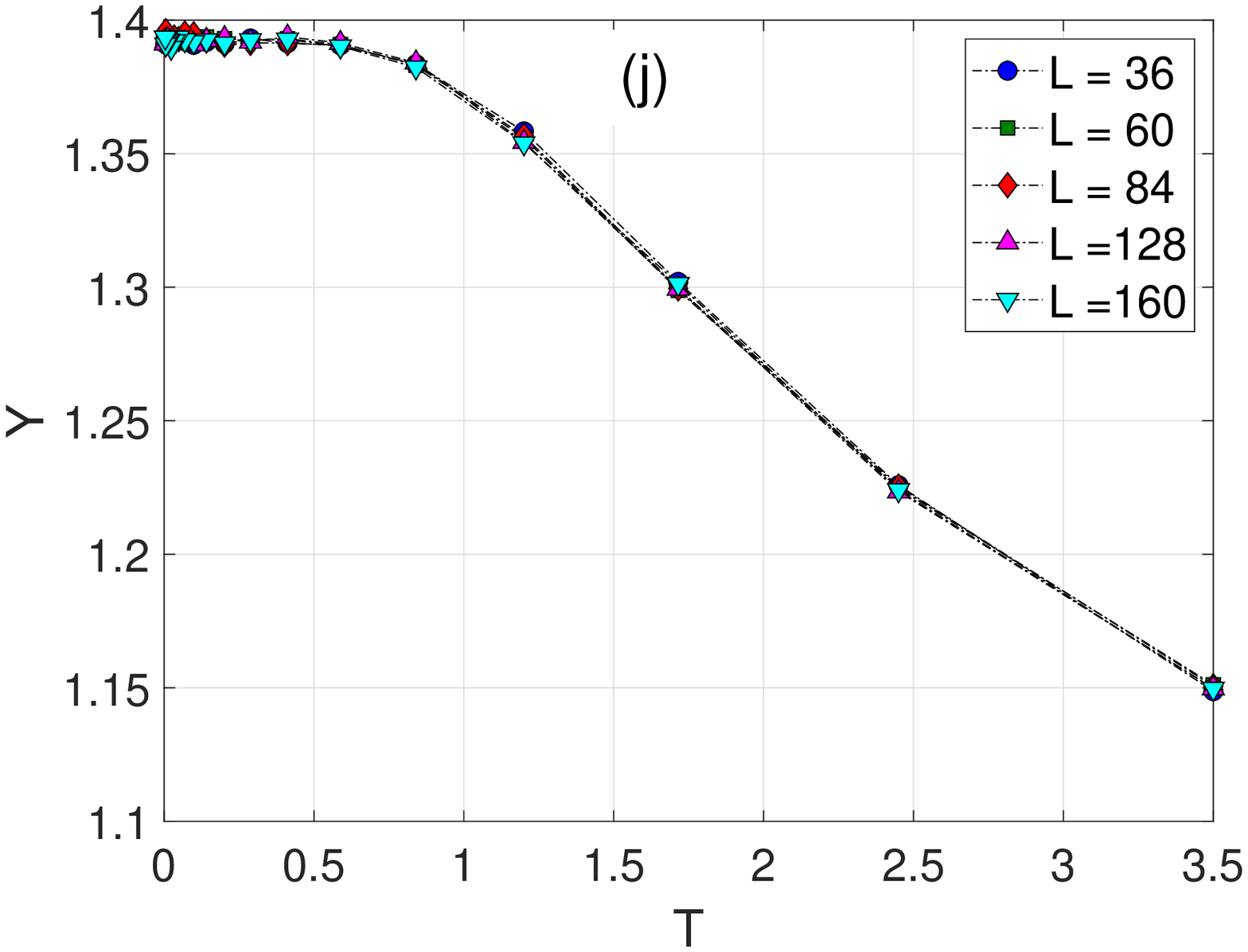}\label{fig:y_s_0_5}}\hspace{-5mm}
    \subfigure{\includegraphics[width = 0.35\textwidth]{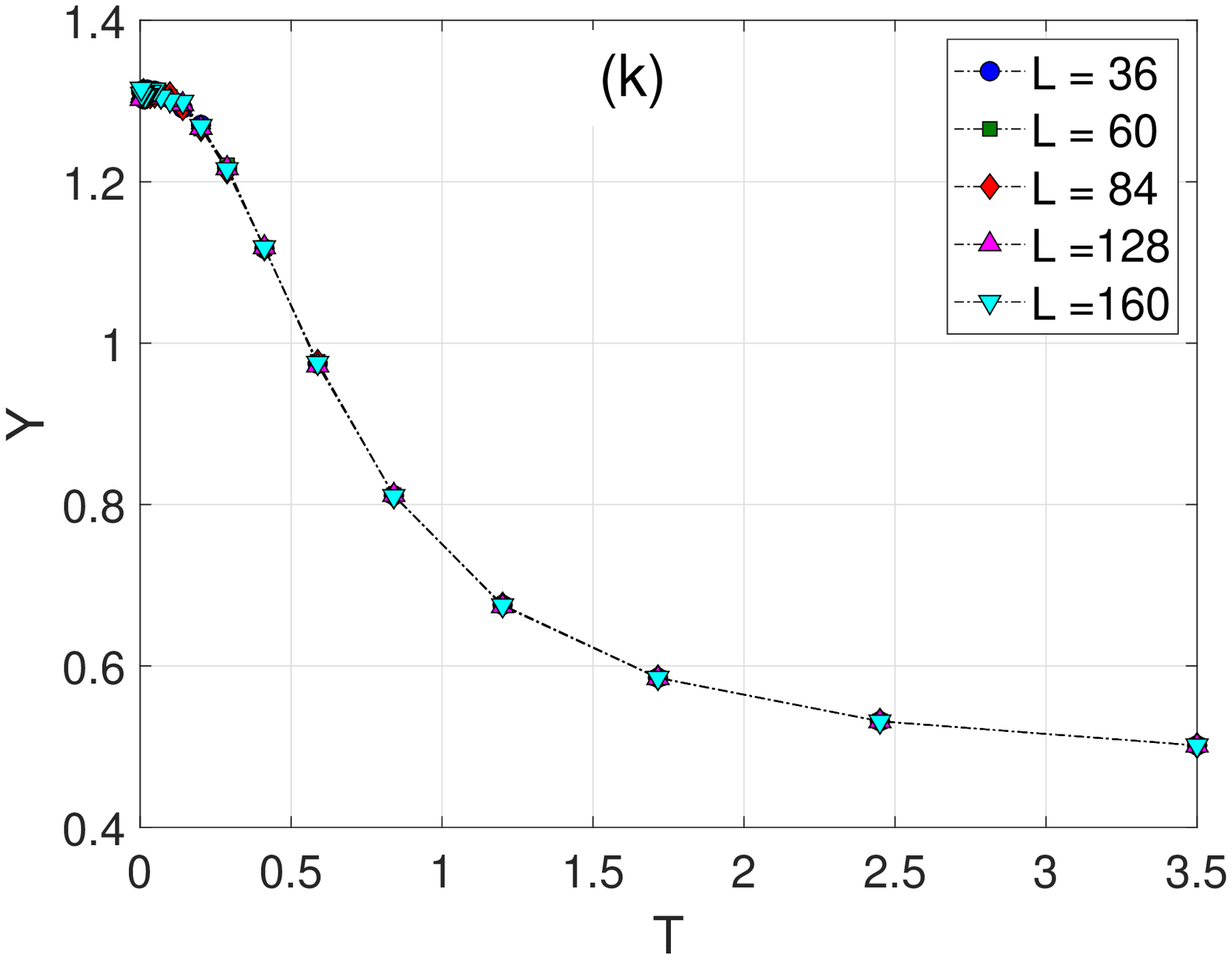}\label{fig:y_s_2_5}}\hspace{-5mm}
		\subfigure{\includegraphics[width = 0.35\textwidth]{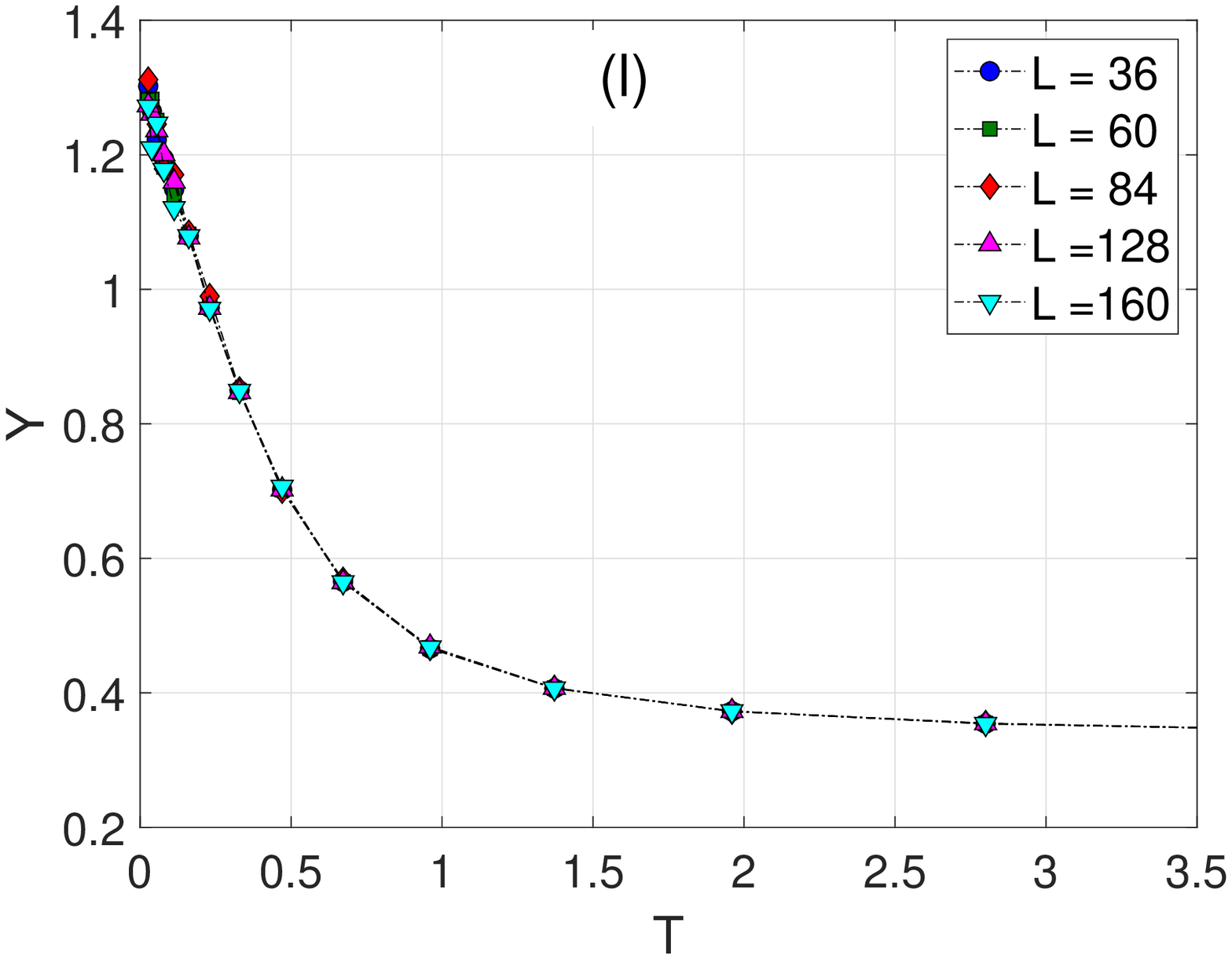}\label{fig:y_s_inf}}
    \caption{Temperature dependencies of (a-c) the internal energy $e$, (d-f) the the specific heat $c$, (g-i) the sublattice magnetizations $m_{\rm A}$, and (j-l) the quantity $Y$, for different $L$ and selected values of $S=1/2$ (left column), $S=5/2$ (central column), and $S=\infty$ (right column).}
    \label{fig:ma_y-T}
\end{figure}

Considering the fact that MC simulations of such a highly frustrated system at very low temperatures are targeted, it is mandatory to ensure that equilibrium conditions are established even for the most demanding situations corresponding to the lowest simulated temperatures and the largest system sizes. Besides some standard analysis of various time series collected during the MC runs, in this case these can also be verified by checking whether the normalized internal energy converges to the expected exact ground-state value of $e_{GS}=-2/3$. 

Figures~\ref{fig:e_s_0_5}-\ref{fig:e_s_inf} present temperature dependencies of $e$ for selected values of $S$, shown on a semi-logarithmic scale for a better focus on the approach to $T \to 0$. A closer inspection reveals that for $S=1/2$ the value of $e_{GS}$ is reached already at relatively high $T \approx 0.3$ (Fig.~\ref{fig:e_s_0_5}), while for $S=5/2$ in order to achieve it the temperature needs to be decreased below $T \approx 0.03$ (Fig.~\ref{fig:e_s_2_5}). With the increasing $S$ the convergence becomes gradually slower and, eventually, for $S=\infty$ the $e_{GS}$ value is not reached even at the lowest simulated temperature $T_{min} = 0.003$ (Fig.~\ref{fig:e_s_inf}). However, the inability to reach the ground-state energy value for $S=\infty$ should not be ascribed to the used method but rather to the high sensitivity of the system to thermal fluctuation that increase the internal energy already at very low temperatures. Equilibrium conditions have been verified and confirmed also in this case by careful analysis of the collected time series. 

Generally, the internal energy per spin is not expected to show any noticeable dependence on the system size and indeed the curves for various $L$ perfectly collapse on each other. The respective specific heat temperature dependencies, plotted in Figs.~\ref{fig:c_s_0_5}-\ref{fig:c_s_inf}, show the presence of a single round maximum moving to lower temperatures with increasing $S$. As expected, for the continuous spin case the specific heat remains finite in the zero temperature limit. Even in this case no finite-size dependence is observed.

In Figs.~\ref{fig:ma_s_0_5}-\ref{fig:ma_s_inf} we show temperature dependencies of the sublattice magnetizations $m_{\rm A}$\footnote{The remaining sublattice magnetizations $m_{\rm B}$ and $m_{\rm C}$ give the same results as $m_{\rm A}$ and, therefore, hereafter only the results for $m_{\rm A}$ and those based on $m_{\rm A}$ will be presented.} and in Figs.~\ref{fig:y_s_0_5}-\ref{fig:y_s_inf} also the quantity $Y$, for some selected values of $S$ and increasing lattice size $L$. While the sublattice magnetizations display overall rather small values, which systematically decrease with the increasing $L$, the values of $Y$ are much larger and do not show any noticeable changes with $L$. As the spin value increases both quantities show a more rapid decrease with the increasing temperature. It is worth mentioning that small values of the sublattice magnetizations close to $T=0$ alone do not rule out the possibility of the sublattice LRO. Unsaturated, albeit much larger ground-state values of $m_{\alpha}$ were also reported in the IATL model for all spins producing sublattice LRO, i.e. $S>S_C$, except for $S=\infty$ (see, e.g., Ref.~\cite{netz93}). We note that in all the curves in Fig.~\ref{fig:ma_y-T}, showing temperature dependencies of various quantities, the error bars are rather small and smaller than the symbol sizes.

\begin{figure}[t!]
		\subfigure{\includegraphics[width = 0.35\textwidth]{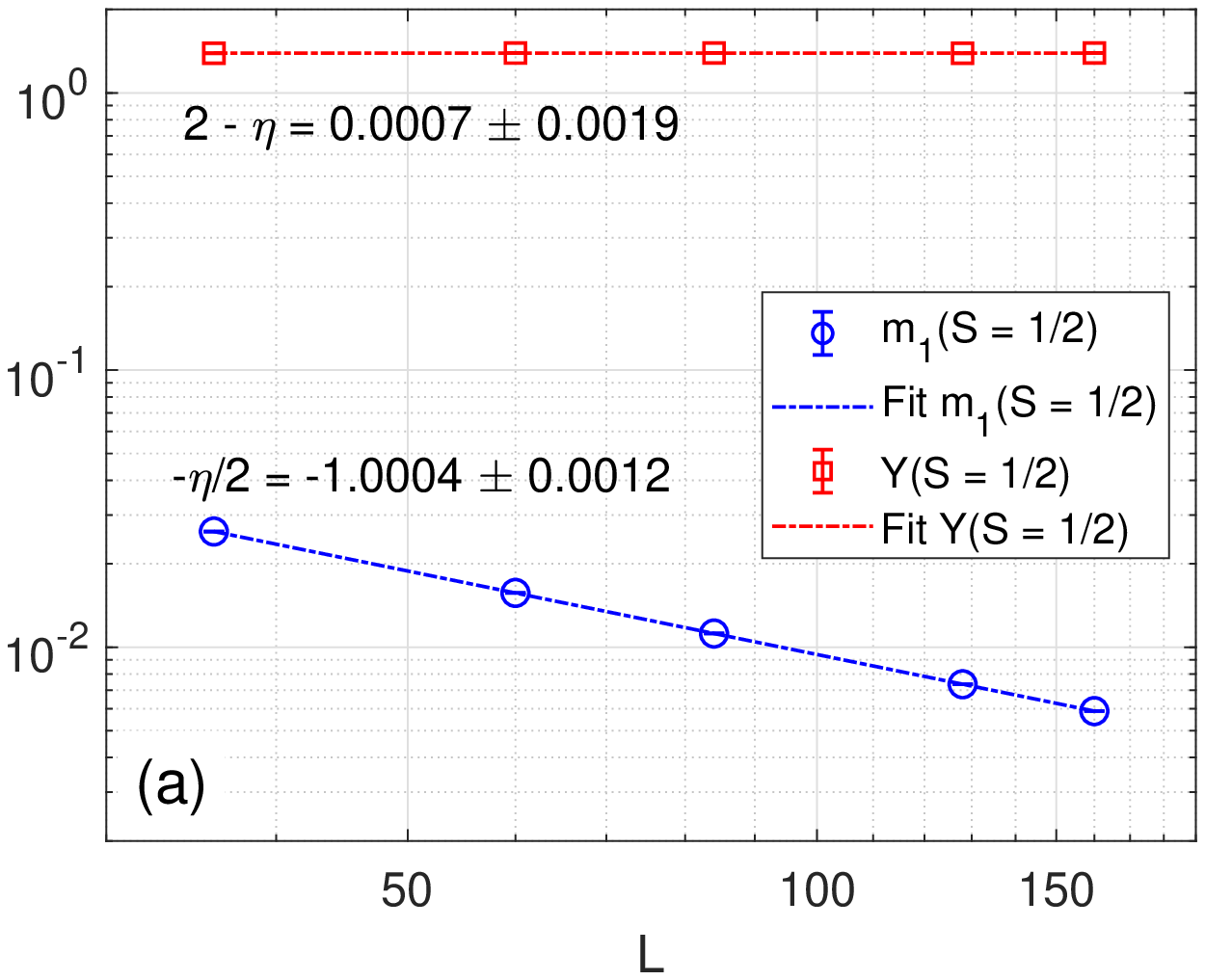}\label{fig:fss_S_0_5_T_min}}\hspace{-5mm}
    \subfigure{\includegraphics[width = 0.35\textwidth]{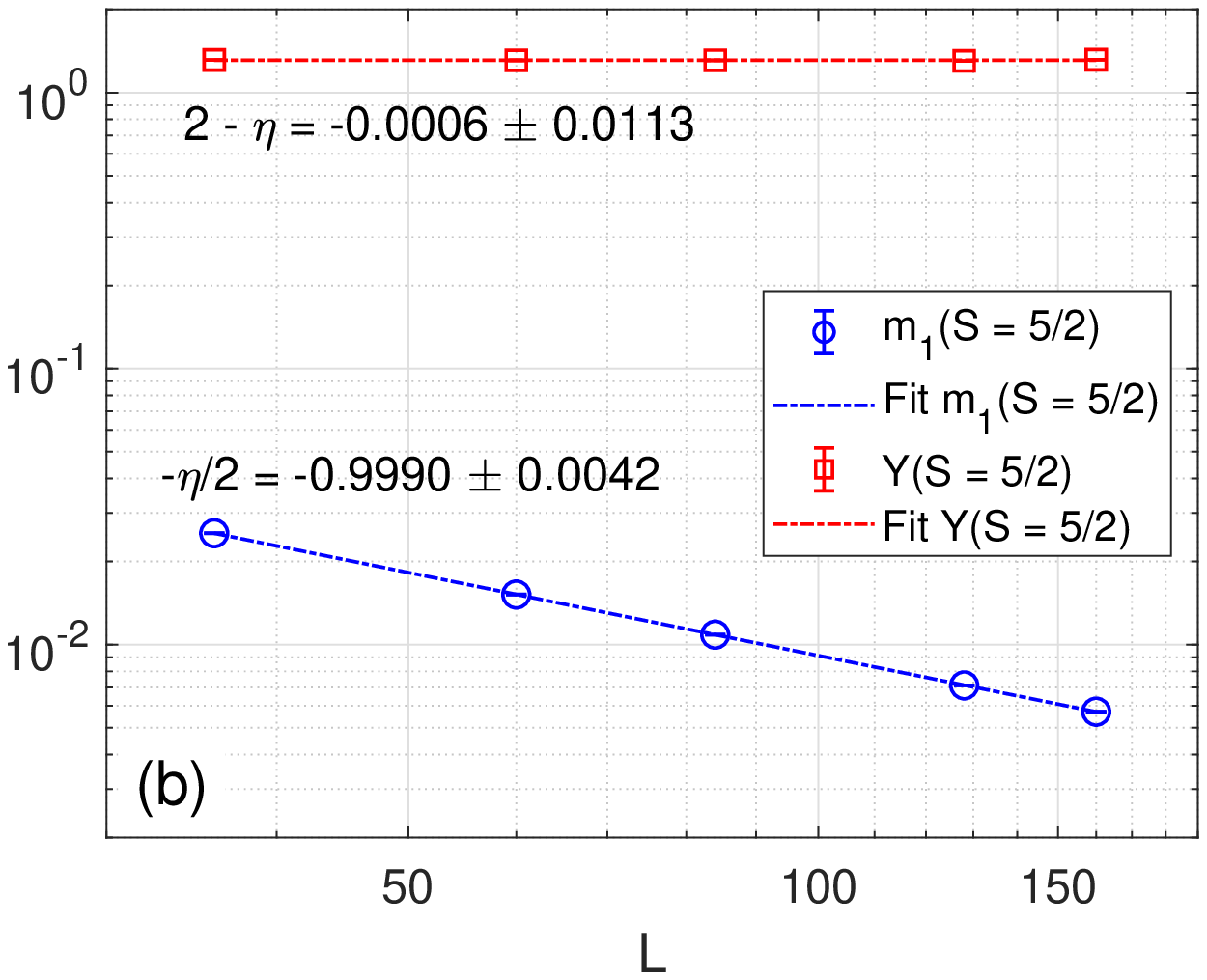}\label{fig:fss_S_2_5_T_min}}\hspace{-5mm}
		\subfigure{\includegraphics[width = 0.35\textwidth]{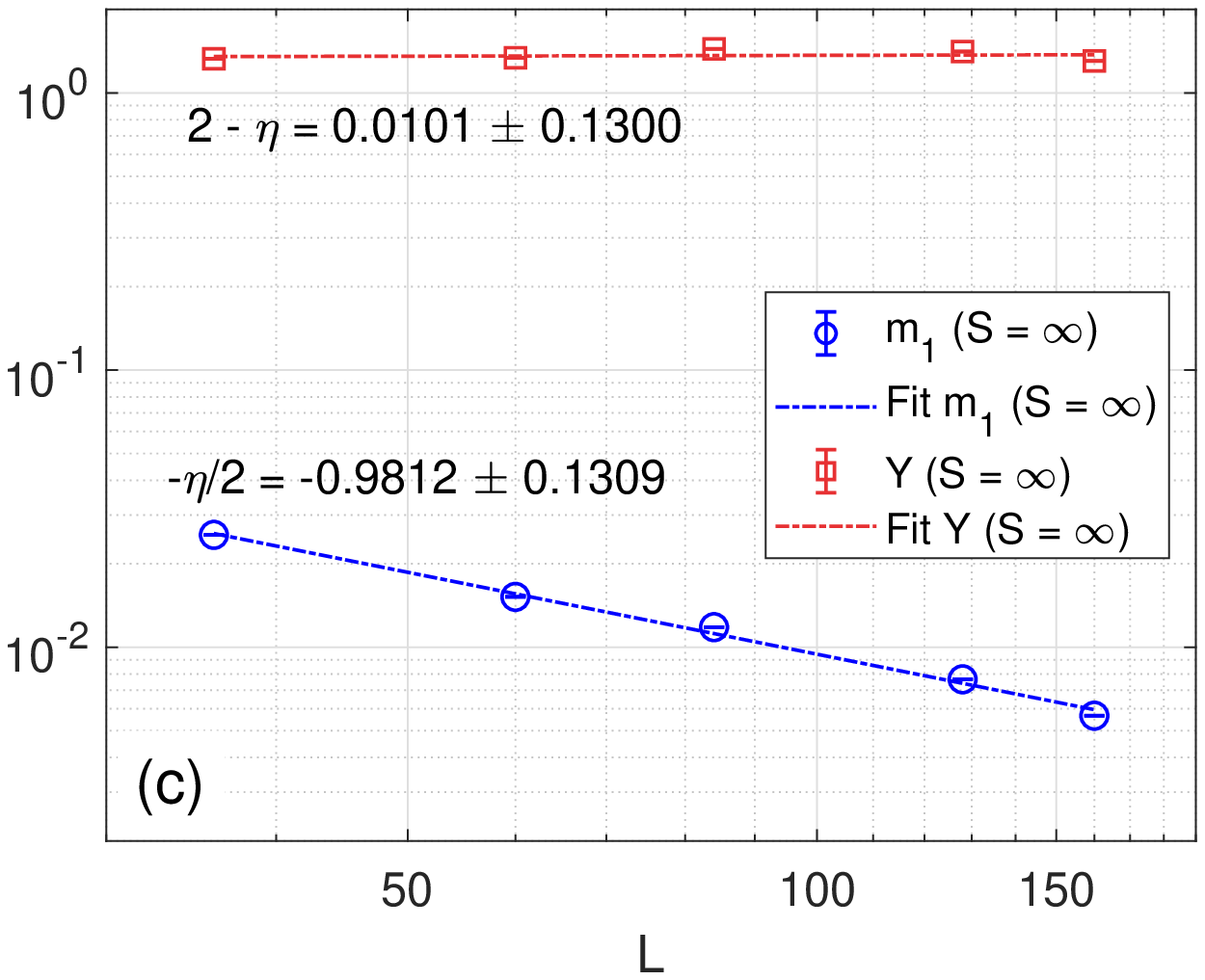}\label{fig:fss_S_inf_T_min}}
    \caption{FSS of the sublattice magnetization $m_{\rm A}$ and the quantity $Y$, for (a) $S=1/2$, (b) $S=5/2$, and (c) $S=\infty$, at $T=0.003$.}
    \label{fig:FSS}
\end{figure}

A better picture about the sublattice magnetizations behavior on approach to the thermodynamic limit can be obtained from a finite-size scaling (FSS) analysis. Using the scaling relations~(\ref{eq:mi_FSS}) and~(\ref{eq:Y_FSS}), in Fig.~\ref{fig:FSS} we present the FSS plots for the sublattice magnetization $m_{\rm A}$ and the quantity $Y$ for different values of $S$ at the lowest considered temperature. All the curves show on the log-log scale an excellent linear dependence. The quality of the fit slightly deteriorates with the increasing value of $S$, nevertheless, even for the worst case of $S=\infty$ an adjusted coefficient of determination, as a measure of goodness of fit~\cite{theil61}, did not drop below $R^2=0.999$. The respective slopes correspond to the values of $-\eta/2 \approx -1$ and $2-\eta \approx 0$, i.e., very close to the value of $\eta = 2$ expected for the exponential decay of the correlation function.

\begin{figure}[t]
    \subfigure{\includegraphics[width = 0.48\textwidth]{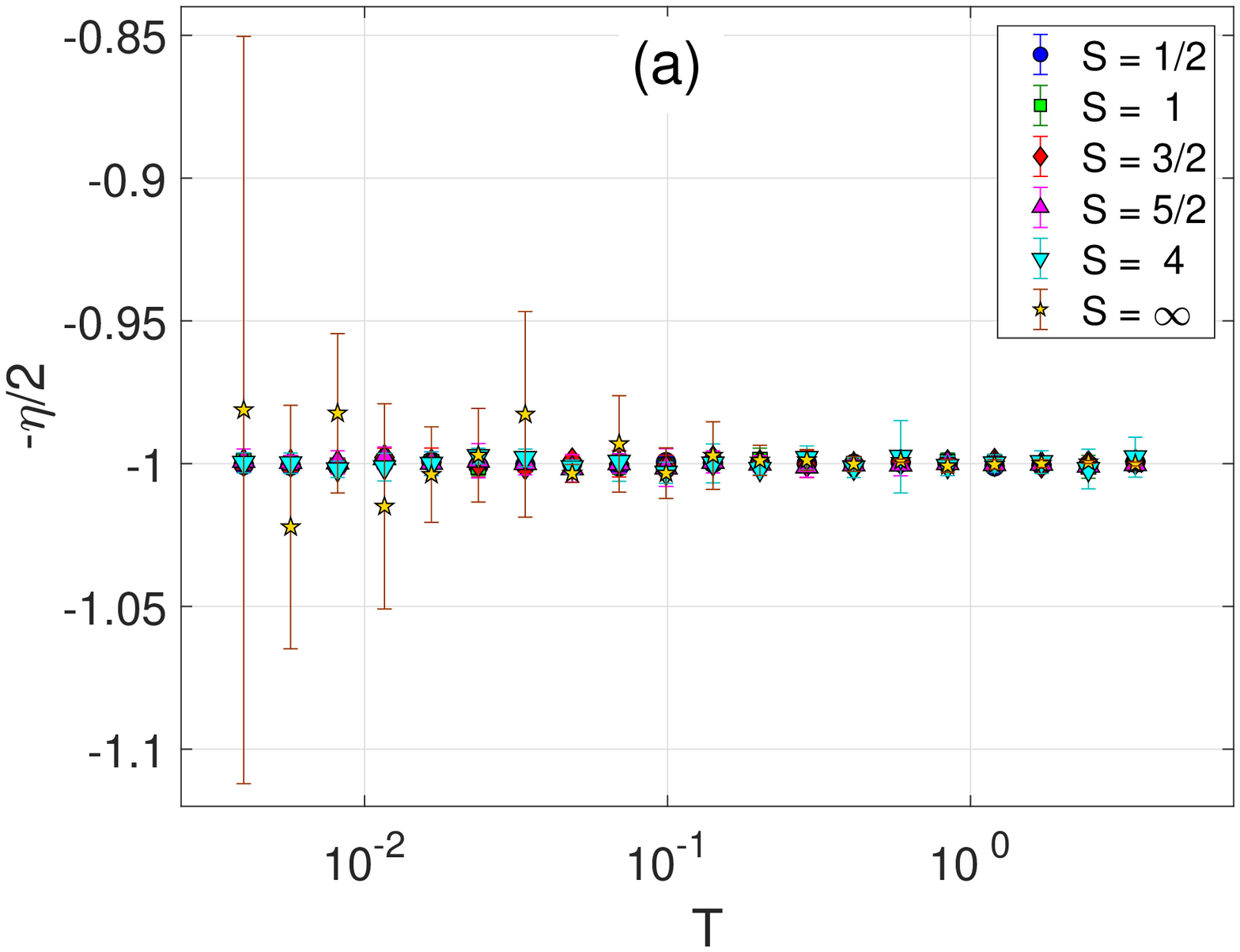}\label{fig:eta_m1-T}}
    \subfigure{\includegraphics[width = 0.48\textwidth]{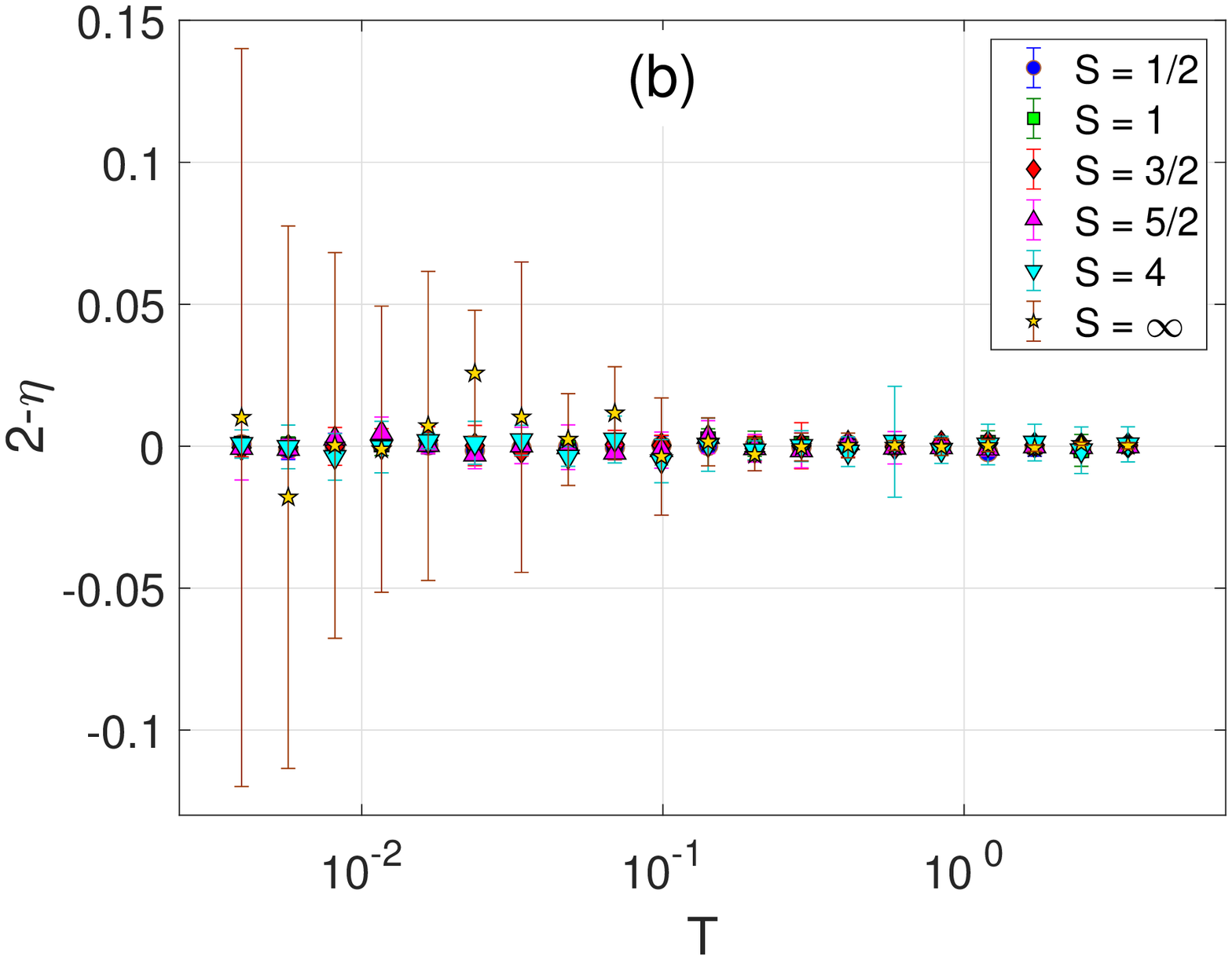}\label{fig:eta_Y-T}}
    \caption{Temperature dependencies of the exponents (a) $-\eta/2$ from FSS of $m_{\rm A}$ and (b) $2-\eta$ from FSS of $Y$, for different values of $S$.}
    \label{fig:T-eta}
\end{figure}

\begin{figure}[t]
    \subfigure{\includegraphics[width = 0.48\textwidth]{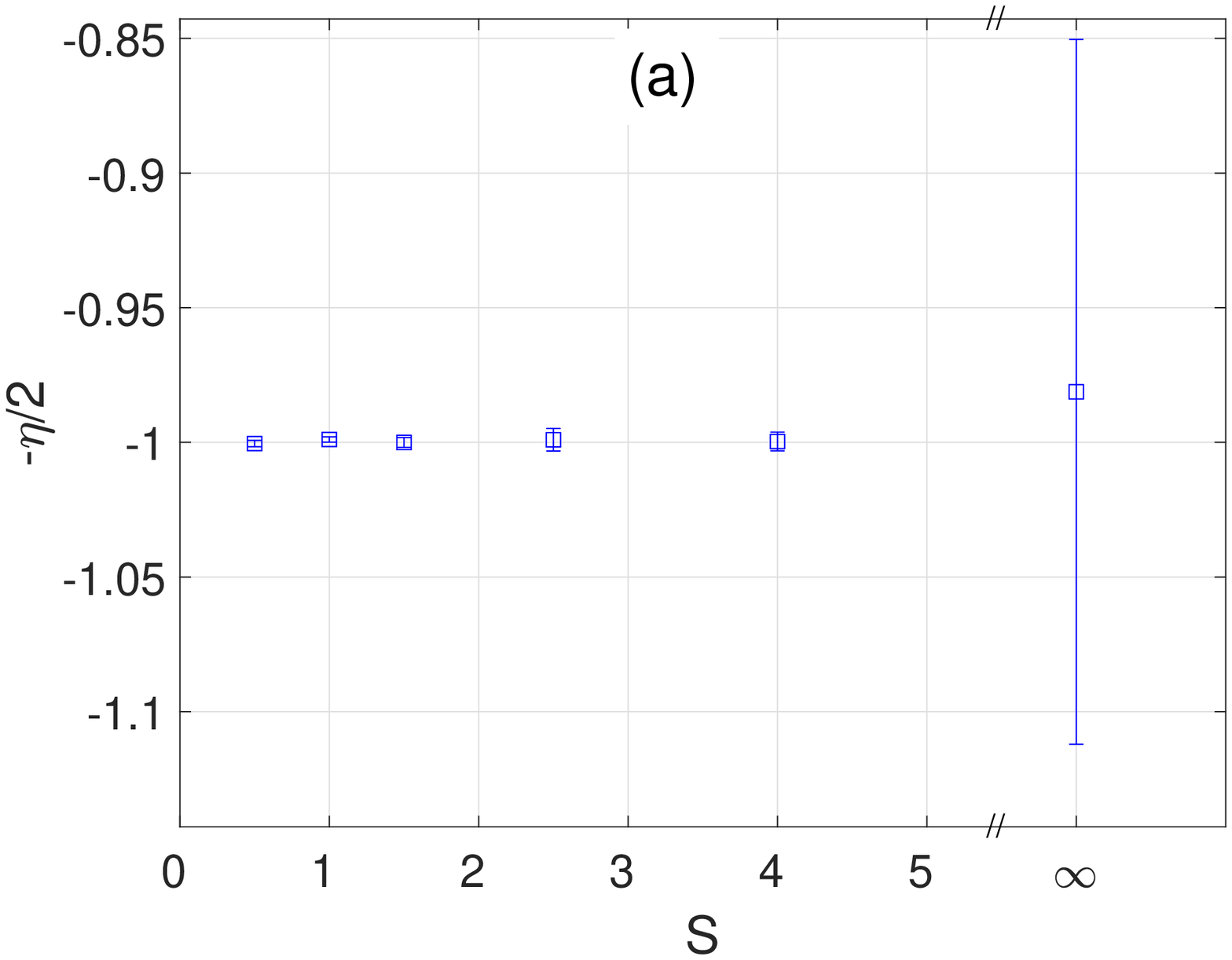}\label{fig:eta_m1-S_T_min}}
    \subfigure{\includegraphics[width = 0.48\textwidth]{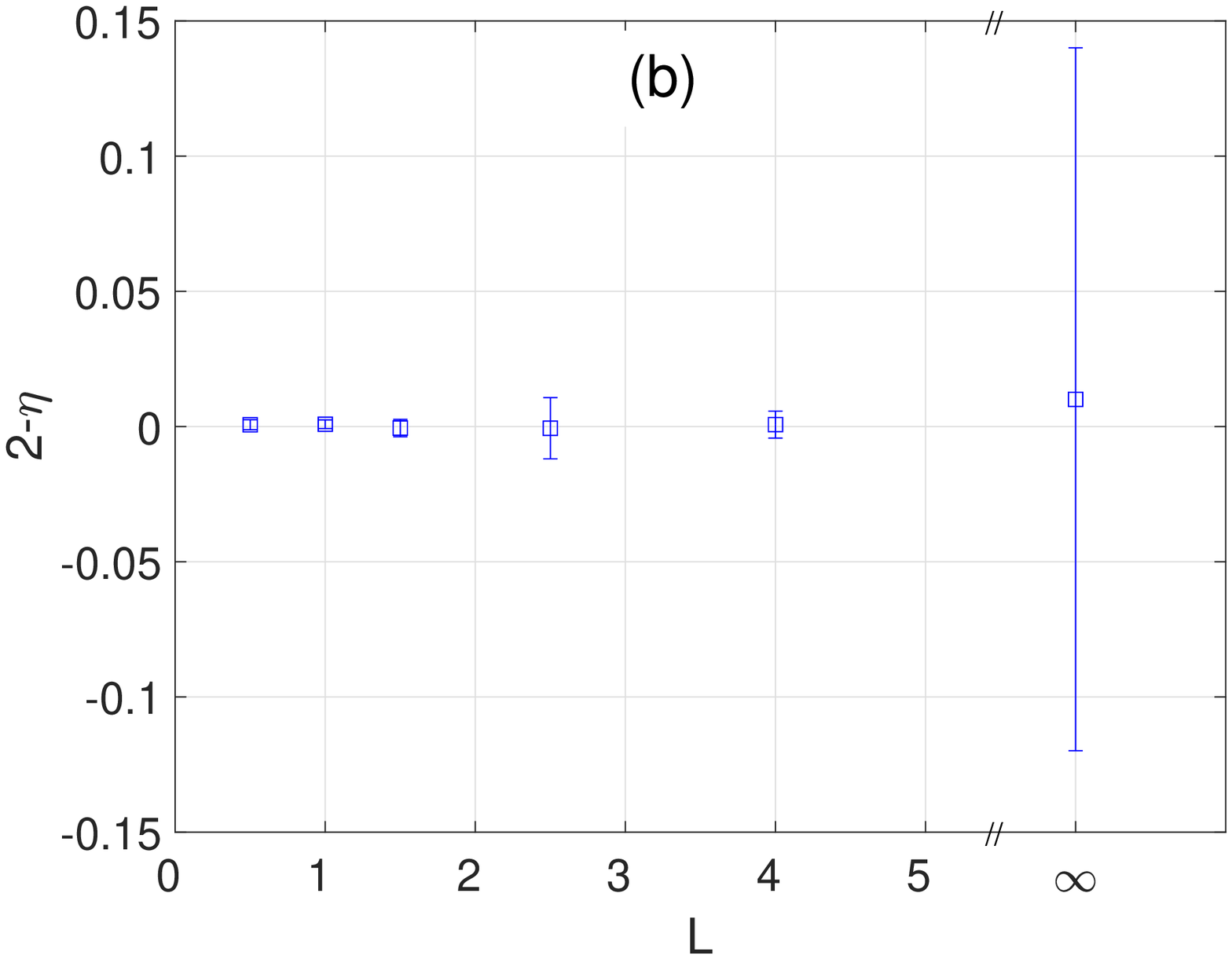}\label{fig:eta_Y-S_T_min}}
    \caption{Spin $S$ dependencies of the exponents (a) $-\eta/2$ from FSS of $m_{\rm A}$ and (b) $2-\eta$ from FSS of $Y$, at $T=0.003$.}
    \label{fig:S-eta}
\end{figure}

Temperature dependencies of the critical exponents $-\eta/2$ and $2-\eta$ for different spin values are shown in Fig.~\ref{fig:T-eta}. For all the temperatures and the spin values the estimated values of $-\eta/2$ and $2-\eta$ fluctuate around $-1$ (Fig.~\ref{fig:eta_m1-T}) and $0$ (Fig.~\ref{fig:eta_Y-T}), respectively. The error bars, which are taken as $95\%$ confidence intervals, are larger at very low $T$ and very large $S$, nevertheless, in all the instances they include the respective values of $-\eta/2=-1$ and $2-\eta=0$. Similar observation can be made by looking at spin $S$ dependencies of the critical exponents $-\eta/2$ and $2-\eta$ at the lowest simulated temperature, shown in Fig.~\ref{fig:S-eta}. The presented evidence strongly suggests that the system with arbitrary spin value $S$ remains in the disordered state down to very low temperatures. 

In this Letter, we studied the possibility of emergence of any long-range ordering (LRO) due to the increase of multiplicity of the local degrees of freedom (spin value $S$) in the IAKL model by the means of Monte Carlo simulations. We considered different values of $S$, including $S=\infty$, and studied the decay of the correlation function by evaluating its critical exponent $\eta$ obtained from a FSS analysis. The goal was to determine whether there exists some threshold value of the spin $S_C$ above which the system would show some LRO, as reported in a related model on triangular lattice (IATL). We found that the value of $\eta$ depends neither on the spin $S$ nor the temperature $T$ and it remains constant at $\eta=2$. This suggests that the IAKL model with general spin $S$ doesn't exhibit any LRO at any finite temperature. Thus this behavior differs from that of the IATL model, for which the system above some $S_C$ crosses over to a partially disordered LRO phase. On the other hand, the finding of no LRO at any finite temperature in the IAKL model resembles more the behavior of its Heisenberg counterpart (HAKL)~\cite{liu16,mull18}, in which, however, LRO is prohibited by the Mermin-Wagner theorem~\cite{merm66}and thus its absence has different origin than in the IAKL model. Nevertheless, the presented finite-temperature results do not exclude the possibility of a crossover to a magnetic LRO phase in the ground state, which is the case for the HAKL system.  

Finally, we would like to remark that the presented analysis focuses on the possibility of one type of sublattice ordering, as shown in Fig. 1 (so-called ``${\mathbf q}=0$''), but in principle it does not rule out some other type of ordering for different choice of sublattices. In fact, we also tentatively performed similar analysis restricted to the case of $S = \infty$ (the most likely candidate for LRO), assuming different, the so-called ``$\sqrt{3}\times\sqrt{3}$'', type of ordering (see, e.g. Fig. 1 in Ref.~\cite{cher14}) but no evidence of LRO was found either. We believe that if there was some type of LRO in the present systems the phase transition to such a state would be accompanied with typical anomalies in the global (involving the entire system) quantities, such as the specific heat and the quantity $Y$, and the type of ordering would be identifiable from the spin snapshots. None of this could be observed in our results and, thus, we find any type of LRO unlikely to occur at least at finite temperatures.

\section*{Acknowledgment}
This work was supported by the Scientific Grant Agency of Ministry of Education of Slovak Republic (Grant No. 1/0531/19), the Slovak Research and Development Agency (Contract No. APVV-16-0186), and the Internal Scientific Grant System of Faculty of Science of UPJ\v{S} (VVGS-2019-1053).


\begin{thebibliography}{50}

\bibitem{wann50} G.H. Wannier, Phys. Rev. 79 (1950) 357.
\bibitem{wann73} G.H. Wannier, Phys. Rev. B 7 (1973) 5017.
\bibitem{hout50} M. Houtappel, Physica 16 (1950) 425.
\bibitem{step70} J. Stephenson, J. Math. Phys. 11 (1970) 413.
\bibitem{hori91} T. Horiguchi, O. Nagai, S. Miyashita, J. Phys. Soc. Jpn. 60 (1991) 1513.    
\bibitem{hori92} T. Horiguchi, O. Nagai, S. Miyashita, Y. Miyatake, Y. Seo, J. Phys. Soc. Jpn. 61 (1992) 3114. 
\bibitem{hori93} T. Horiguchi, O. Nagai, H.T. Diep, Y. Miyatake, Phys. Lett. A 177 (1993) 93.
\bibitem{naga93} O. Nagai, S. Miyashita, T. Horiguchi, Phys. Rev. B 47 (1993) 202.
\bibitem{netz93} R.R. Netz, Phys. Rev. B 48 (1993) 16113.
\bibitem{naga94} O. Nagai, M. Kang, Y. Yamada, T. Horiguchi, Phys. Lett. A 195 (1994) 263.
\bibitem{yama95} Y. Yamada, S. Miyashita, T. Horiguchi, M. Kang, O. Nagai, J. Magn. Magn. Mater. 140 (1995) 1749.
\bibitem{lipo95} A. Lipowski, T. Horiguchi, D. Lipowska, Phys. Rev. Lett. 74 (1995) 3888.
\bibitem{zeng97} C. Zeng, C.L. Henley, Phys. Rev. B 55 (1997) 14935.
\bibitem{ande73} P.W. Anderson, Mater. Res. Bull. 8 (1973) 153.
\bibitem{faze74} P. Fazekas, P.W. Anderson, Philos. Mag. 30 (1974) 423.
\bibitem{joli89} T. Jolicoeur, J.C. LeGouillou, Phys. Rev. B 40 (1989) 2727.
\bibitem{bern92} B. Bernu, C. Lhuillier, L. Pierre, Phys. Rev. Lett. 69 (1992) 2590.
\bibitem{chub94} A.V. Chubukov, S. Sachdev, T. Senthil, J. Phys.: Condens. Matter 6 (1994) 8891.
\bibitem{capr99} L. Capriotti, A.E. Trumper, S. Sorella, Phys. Rev. Lett. 82 (1999) 3899.
\bibitem{whit07} S.R. White, A.L. Chernyshev, Phys. Rev. Lett. 99 (2007) 127004.
\bibitem{zhit13} M.E. Zhitomirsky, A.L. Chernyshev, Rev. Mod. Phys. 85 (2013) 219.
\bibitem{gotz16} O. G\"{o}tze, J. Richter, R. Zinke, D.J.J. Farnell, J. Magn. Magn. Mater. 397 (2016) 333.
\bibitem{sach92} S. Sachdev, Phys. Rev. B 45 (1992) 12377.
\bibitem{chub92} A. Chubukov, Phys. Rev. Lett. 69 (1992) 832.
\bibitem{henl95} C.L. Henley, E.P. Chan, J. Magn. Magn. Mater. 140 (1995) 1693.
\bibitem{yan11} S. Yan, D.A. Huse, S.R. White, Science 332 (2011) 1173.
\bibitem{gotz11} O. G\"{o}tze, D.J.J. Farnell, R.F. Bishop, P.H.Y. Li, J. Richter, Phys. Rev. B 84 (2011) 224428.
\bibitem{depe12} S. Depenbrock, I.P. McCulloch, U. Schollwock, Phys. Rev.
Lett. 109 (2012) 067201. 
\bibitem{jian12} H.C. Jiang, Z.H. Wang, L. Balents, Nat. Phys. 8 (2012) 902.
\bibitem{iqba15} Y. Iqbal, D. Poilblanc, F. Becca, Phys. Rev. B 91 (2015) 020402(R).
\bibitem{liu15} T. Liu, W. Li, A. Weichselbaum, J. von Delft, G. Su, Phys. Rev. B 91 (2015) 060403.
\bibitem{chan15} H.J. Changlani, A.M. L\"{a}uchli, Phys. Rev. B 91 (2015) 100407.
\bibitem{nish15} S. Nishimoto, M. Nakamura, Phys. Rev. B 92 (2015) 140412(R).
\bibitem{li15} W. Li, A. Weichselbaum, J. von Delft, H.-H. Tu, Phys. Rev. B 91 (2015) 224414.
\bibitem{cher14} A.L. Chernyshev, M.E. Zhitomirsky, Phys. Rev. Lett. 113 (2014) 237202.
\bibitem{gotz15} O. G\"{o}tze, J. Richter, Phys. Rev. B 91 (2015) 104402.
\bibitem{oitm16} J. Oitmaa, R.R.P. Singh, Phys. Rev. B 93 (2016) 014424.
\bibitem{liu16} T. Liu, W. Li, G. Su, Phys. Rev. B 94 (2016) 032114.
\bibitem{mull18} P. M\"{u}ller, A. Zander, J. Richter, Phys. Rev. B 98 (2018) 2.
\bibitem{syoz51} I. Syozi, Prog. Theor. Phys. 6 (1951) 306.
\bibitem{kano53} K. Kano, S. Naya, Progr. Theor. Phys. 10 (1953) 158.
\bibitem{harr92} A.B. Harris, C. Kallin, A.J. Berlinsky, Phys. Rev. B 45 (1992) 2899.
\bibitem{chalk92} J.T. Chalker, P.C.W. Holdsworth, E.F. Shender, Phys. Rev. Lett. 68 (1992) 855.
\bibitem{reim93} J.N. Reimers, A.J. Berlinsky, Phys. Rev. B 48 (1993) 9539.
\bibitem{delf93} J. von Delft, C.L. Henley, Phys. Rev. B 48 (1993) 965.
\bibitem{chal86} M.S.S. Challa, D.P. Landau, Phys. Rev. B 33 (1986) 437.
\bibitem{miya86} S. Miyashita, Prog. Theor. Phys. Suppl. 87 (1986) 112.
\bibitem{theil61} H. Theil, Economic Forecasts and Policy, Vol. XV of Contributions to Economic Analysis (North-Holland, Amsterdam, 1961).
\bibitem{merm66} N.D. Mermin, H. Wagner, Phys. Rev. Lett. 17 (1966) 1133.
\end{thebibliography}
\end{document}